\documentclass[11pt]{article}
\pdfoutput=1
\usepackage{fullpage}
\usepackage{cite}
\usepackage{graphicx}
\usepackage{amsmath}
\usepackage{amssymb}
\usepackage{subcaption}
\usepackage[utf8x]{inputenc} 
\title{}
\date{}
\renewcommand{\vec}[1]{\mbox{\boldmath$ #1 $}}

\def\beq{\begin{equation}}
\def\eeq{\end{equation}}
\allowdisplaybreaks
\begin{document}
\bibliographystyle{utphys}

\newcommand\n[1]{\textcolor{red}{(#1)}} 
\newcommand{\diff}{\mathop{}\!\mathrm{d}}
\newcommand{\lb}{\left}
\newcommand{\rb}{\right}
\newcommand{\f}{\frac}
\newcommand{\pd}{\partial}
\newcommand{\tr}{\text{tr}}
\newcommand{\fdiff}{\mathcal{D}}
\newcommand{\im}{\text{im}}
\let\caron\v
\renewcommand{\v}{\mathbf}
\newcommand{\T}{\tensor}
\newcommand{\R}{\mathbb{R}}
\newcommand{\C}{\mathbb{C}}
\newcommand{\Z}{\mathbb{Z}}
\newcommand{\msbar}{\ensuremath{\overline{\text{MS}}}}
\newcommand{\DIS}{\ensuremath{\text{DIS}}}
\newcommand{\abar}{\ensuremath{\bar{\alpha}_S}}
\newcommand{\bb}{\ensuremath{\bar{\beta}_0}}
\newcommand{\rc}{\ensuremath{r_{\text{cut}}}}
\newcommand{\Nd}{\ensuremath{N_{\text{d.o.f.}}}}
\setlength{\parindent}{0pt}

\titlepage
\begin{flushright}
QMUL-PH-20-09\\
UCLA/TEP/2020/103
\end{flushright}

\vspace*{0.5cm}

\begin{center}
{\bf \Large The convolutional double copy: a case study with a point}

\vspace*{1cm} 
\textsc{Andr\'{e}s Luna$^a$\footnote{luna@physics.ucla.edu},
Silvia Nagy$^b$\footnote{Silvia.Nagy@nottingham.ac.uk},
  and Chris D. White$^c$\footnote{christopher.white@qmul.ac.uk}} \\

\vspace*{0.5cm} $^a$ Mani L. Bhaumik Institute for Theoretical Physics,
UCLA Department of Physics and Astronomy, Los
Angeles, CA 90095, USA\\

\vspace*{0.5cm} $^b$ Centre for Astronomy and Particle Theory, University
Park, \\ Nottingham, NG7 2RD, United Kingdom

\vspace*{0.5cm} $^c$ Centre for Research in String Theory, School of
Physics and Astronomy, \\
Queen Mary University of London, 327 Mile End
Road, London E1 4NS, UK\\

\end{center}

\vspace*{0.5cm}

\begin{abstract}
The double copy relates scattering amplitudes in gauge and gravity
theories. It has also been extended to classical solutions, and a
number of approaches have been developed for doing so. One of these
involves expressing fields in a variety of (super-)gravity theories in
terms of convolutions of gauge fields, including also BRST ghost
degrees of freedom that map neatly to their corresponding counterparts
in gravity. In this paper, we spell out how to use the convolutional
double copy to map gauge and gravity solutions in the manifest Lorenz
and de Donder gauges respectively. We then apply this to a particular
example, namely the point charge in pure gauge theory. As well as
clarifying how to use the convolutional approach, our results provide
an alternative point of view on a recent discussion concerning whether
point charges map to the Schwarzschild solution, or the more general
two-parameter JNW solution, which includes a dilaton field. We confirm
the latter.
\end{abstract}

\vspace*{0.5cm}

\section{Introduction}
\label{sec:intro}

The BCJ double copy of
refs.~\cite{Bern:2008qj,Bern:2010ue,Bern:2010yg} is by now a
well-established relationship between scattering amplitudes in gauge
and gravity theories. Since its inception, a number of approaches have
tried to extend its remit to classical solutions, exact or
otherwise. Examples include the use of Kerr-Schild
coordinates~\cite{Monteiro:2014cda,Luna:2015paa,Luna:2016due,Bahjat-Abbas:2017htu,Berman:2018hwd,Ridgway:2015fdl,Carrillo-Gonzalez:2017iyj,CarrilloGonzalez:2019gof,Bahjat-Abbas:2020cyb,Alfonsi:2020lub,Alawadhi:2019urr},
spinorial methods~\cite{Monteiro:2018xev,Luna:2018dpt}, worldline
methods~\cite{Goldberger:2016iau,Goldberger:2017frp,Goldberger:2017vcg,Goldberger:2017ogt,Goldberger:2019xef},
perturbative diagrammatic
reasoning~\cite{Luna:2016hge,Luna:2017dtq,Maybee:2019jus,Borsten_Nagy},
and double field theory~\cite{Lee:2018gxc,Kim:2019jwm}. In this paper,
we will focus on another approach, first introduced in
ref.~\cite{Anastasiou:2014qba}, in which the field content of a
gravity theory in position space can be obtained using convolutions of
fields from a gauge theory. The convolution operation is defined for
two functions $f(x)$ and $g(x)$ via
\begin{equation}
[f\star g](x)=\int d^4y f(x) g(x-y).
\label{fstarg}
\end{equation}
This is both commutative and associative, and furthermore obeys the
derivative rule
\begin{equation}
\partial_\mu(f\star g)=(\partial_\mu f)\star g=f\star (\partial_\mu g).
\label{derivrule}
\end{equation}
The motivation for this product is that, for scattering amplitudes,
the double copy operates via products of functions in momentum
space. Upon Fourier transforming back to position space, such products
would become the convolutions of eq.~(\ref{fstarg}). \\

A restriction of the convolution approach is that it is currently only
defined for linearised gauge and gravity theories. However, the
Kerr-Schild double copy of ref.~\cite{Monteiro:2014cda} also relates
solutions of the linear equations (which happen in that case to be
solutions of the full non-linear theory). Furthermore, working at
linear level for more general solutions certainly does not prevent the
use of the convolution formalism to gain significant insights. It was
used in
refs.~\cite{Anastasiou:2016csv,Cardoso:2016ngt,Cardoso:2016amd,Anastasiou:2017nsz,Anastasiou:2017taf,Borsten:2013bp,Anastasiou:2013hba,Anastasiou:2015vba},
for example, to construct a wide catalogue of double copy examples,
involving highly exotic (super-)gravity theories. Another significant
advantage of the convolution approach is that it works, in principle,
for arbitrary gauge choices in both the gauge and gravity theories, a
feature which is not typically shared by other classical double copy
approaches, or by the original BCJ double copy for amplitudes. The
latter relies on a certain duality between colour and kinematics being
made manifest~\cite{Bern:2008qj}, which is possible in principle using
both gauge transformations and field redefinitions~\footnote{See,
  however, ref.~\cite{Bern:2017yxu,Bern:2017ucb}, wich describes how
  the BCJ duality requirement can be relaxed.}, collectively known as
{\it generalised gauge transformations}.\\

In a non-abelian gauge theory, a gauge-fixing condition gives rise to
ghost fields according to the usual Faddeev-Popov
procedure~\cite{Faddeev:1967fc}, and the resulting action obeys the
well-known BRST
invariance~\cite{Becchi:1974xu,Becchi:1974md,Becchi:1975nq,Tyutin:1975qk},
where an analogous story holds in
(super-)gravity~\cite{Fradkin:1975cq,Batalin:1977pb}. How to marry the
BRST formalism with the double copy was addressed in
refs.~\cite{Anastasiou:2018rdx,Borsten:2019prq}, which extended the
convolutional double copy to include ghost fields, such that all
physical degrees of freedom in the gravity theory can be obtained by
combining both physical and ghost fields from the two chosen gauge
theories (see also refs.~\cite{Siegel:1988qu,Siegel:1995px} for
earlier work describing such a correspondence). It was also used to
study the issue of the separation of the dilaton degree of freedom
from the trace of the graviton, pointed out in e.g.
refs.~\cite{LopesCardoso:2018xes,Luna:2016hge}.  The particular
dictionary between gauge and gravity fields is not unique, but will
depend on the choices of gauge-fixing condition on both the gauge and
gravity sides. Although the relevant principles were defined in
ref.~\cite{Anastasiou:2018rdx}, there have to date been few results in
the literature regarding both (i) how to construct the convolutional
double copy dictionary
explicitly~\footnote{Reference~\cite{Borsten:2020bgv} provides a
  recent pedagogical review of this and many other double-copy-related
  topics.}, and (ii) its consequences for particular solutions. The
aim of this paper is to rectify this situation, by carrying out the
convolutional double copy for particular classical solutions, with
particular gauge choices. \\

More specifically, we will examine the expression for a point charge
in pure (non-supersymmetric) Yang-Mills theory, and construct its
double copy in the relevant gravity theory, namely ${\cal N}=0$
Supergravity. The latter consists of General Relativity coupled to a
dilaton and axion (two-form) field, and we will choose the
commonly-used Lorenz and de Donder gauges in the gauge and gravity
theory respectively. There are good reasons to choose this particular
example. Firstly, it is arguably the simplest case of a gauge theory
solution that one may use to probe the classical double
copy. Secondly, there has been some discussion in the recent
literature regarding the precise identity of the point charge's double
copy. The Kerr-Schild approach of ref.~\cite{Monteiro:2014cda} was the
first to consider this, and identified the point charge with a
Schwarzschild black hole. However, ref.~\cite{Goldberger:2016iau} used
worldline methods to point out that one typically expects the dilaton
field to be turned on in the gravity theory, which is seemingly at
odds with ref.~\cite{Monteiro:2014cda}~\footnote{Interestingly, the
  dilaton turns on in three spacetime dimensions, even in the
  Kerr-Schild
  approach~\cite{CarrilloGonzalez:2019gof}.}. Reference~\cite{Luna:2016hge}
argued that indeed the most general double copy of the point charge
should be regarded as the two-parameter JNW
solution~\cite{Janis:1968zz}, consisting of a spherically symmetric
graviton and dilaton system. By choosing generalised gauges
appropriately, one may select any special case of JNW, including
Schwarzschild. This point of view was corroborated by the recent study
of ref.~\cite{Kim:2019jwm}, using double field theory. We will see
that the BRST convolutional double copy provides a highly useful
complementary analysis of this situation. We will demonstrate
explicitly how the two-parameter JNW solution is obtained, whilst also
seeing how the various ghost degrees of freedom can be independently
chosen so as to restrict to particular cases. Our results thus clarify
the convolutional double copy framework, whilst also exhibiting a
topical application.\\

The structure of our paper is as follows. In section~\ref{sec:local},
we describe the convolutional double copy in more detail, and try to
construct the simplest possible dictionary between gauge and gravity
fields in the Lorenz and de Donder gauges respectively. In this
warm-up example, we will omit possible terms containing inverse
derivatives, which correspond to non-local operators in position
space. However, we will see that this prescription is insufficient to
obtain the full two parameter JNW solution in de Donder gauge. Thus, in
section~\ref{sec:nonlocal}, we will append the dictionary accordingly,
showing explicitly how the point charge and JNW solutions are
related. Finally, we discuss our results and conclude in
section~\ref{sec:conclude}.

\section{From Lorenz to de Donder: a local dictionary}
\label{sec:local}

In this section, we give our first illustration of the convolutional
double copy, by showing how one may construct a dictionary between
Yang-Mills and gravity fields, including ghosts (see also
ref.~\cite{Anastasiou:2018rdx,Borsten:2019prq}). As discussed above, we will first
consider only terms containing products of fields, without including
inverse derivatives. This will act as a practice run for a more
general approach, to be considered in what follows. Also, although the
details of this dictionary have appeared before in the literature, it
has not been applied to the point charge. We will see below that this
has interesting consequences.\\

We will focus on the case in which both gauge theories entering the
double copy are the same. One may then define the {\it circle product}
of two non-abelian gauge fields~\cite{Anastasiou:2014qba}:
\begin{equation}
A_\mu\circ A_\nu\equiv A_\mu^a\star \Phi^{aa'}\star A_\nu^{a'},
\label{circprod}
\end{equation}
where $\Phi^{aa'}$ is a so-called {\it spectator field}, that couples
together the adjoint indices $(a,a')$. This is a scalar field with two
adjoint indices, and is related to the biadjoint scalar field that
appears in other classical double copy
approaches~\cite{Monteiro:2014cda,White:2016jzc,DeSmet:2017rve,Bahjat-Abbas:2018vgo}~\footnote{The
  field $\Phi^{aa'}$ as defined here is actually the convolution {\it
    inverse} of the biadjoint scalar field, and corrects for the fact
  that the biadjoint field appears in $A_\mu^a$, and thus has been
  counted twice in eq.~(\ref{circprod}).}. A similar product may be
used for the (anti)-ghost fields $(c^a,\bar{c}^a)$ appearing in the
Yang-Mills theory, where from now on we will suppress adjoint indices
for brevity, unless otherwise stated. \\

The first step in applying the convolutional double copy is to write a
general ansatz expressing the graviton $h_{\mu\nu}$ and dilaton field
$\phi$ in terms of the Yang-Mills fields
$(A_\mu,c,\bar{c})$~\footnote{The antisymmetric two-form field
  $B_{\mu\nu}$ can only appear if the two gauge fields are chosen to
  be different.}:
\begin{align}
h_{\mu\nu}&=2A_\mu\circ A_\nu+\eta_{\mu\nu}\left(a_1 A^\rho\circ A_\rho
+a_2 c\circ \bar{c}\right); \label{dict1a}\\
\varphi&=a_3A^\rho\circ A_\rho+a_4 c\circ\bar{c}.
\label{dict1b}
\end{align}
Without loss of generality, we have used an overall normalisation
constant on the right-hand side of eq.~(\ref{dict1a}), to fix a factor
of 2 in the first term to match the normalisation for the general
asymmetric case~\cite{Anastasiou:2018rdx}. The Yang-Mills action,
including ghosts, is invariant under the BRST transformations
\begin{equation}
QA_\mu=\partial_\mu c,\quad Qc=0,\quad Q\bar{c}=\frac{1}{\xi}G[A_\mu],
\label{YMBRST}
\end{equation}
where $Q$ is a Grassmann-valued charge,
\begin{equation}
G[A]=\partial\cdot A
\label{Lorenz}
\end{equation}
the gauge-fixing condition, which we have chosen to correspond to the
Lorenz gauge, and $\xi$ an arbitrary multiplier that defines the usual
general family of such covariant gauges. A crucial observation of
ref.~\cite{Anastasiou:2018rdx} is that one should identify the BRST
operator $Q$ with its counterpart in the gravity theory, which acts on
the graviton and dilaton via
\begin{equation}
Qh_{\mu\nu}=\partial_{\mu}c_{\nu}+\partial_\nu c_\mu,\quad
Q\varphi=0.
\label{gravBRST}
\end{equation}
Here $c_\mu$ are the ghosts for the gravity field, and
requiring the BRST constraints to be satisfied in the gravity theory
allows the parameters entering the dictionary of eqs.~(\ref{dict1a},
\ref{dict1b}) to be constrained. To this end, one may use
eq.~(\ref{YMBRST}) and the (anti-)linearity of the BRST charge
\begin{equation}
Q[A\circ B]=(QA)\circ B\pm A\circ (QB),
\label{Qprod}
\end{equation}
where the upper (lower) sign holds if $A$ is bosonic (fermionic), to get
\begin{equation}
Qh_{\mu\nu}=4\partial_{(\mu}c \circ A_{\nu)}+\eta_{\mu\nu}
\left(2a_1-\frac{a_2}{\xi}\right)
\partial^\rho c\circ A_\rho,
\label{dictcalc1}
\end{equation}
where we define
\begin{equation}
a_{(\mu}b_{\nu)}=\frac12 \left(a_\mu b_\nu+a_\nu b_\mu\right)
.
\label{symmetrisation}
\end{equation}
Consistency with eq.~(\ref{gravBRST}) immediately gives the constraint
\begin{equation}
a_2=2\xi a_1.
\label{a2val}
\end{equation}
and then we can read off
\begin{equation}
c_\mu=2c\circ A_{\nu},\quad \Rightarrow \quad
\bar{c}_\mu=2\bar{c}\circ A_{\nu},
\label{cresult}
\end{equation}
where the second condition for the anti-ghost follows from
conjugation. Similarly, eq.~(\ref{dict1b}) yields
\begin{equation}
Q\varphi=\left(2a_3-\frac{a_4}{\xi}\right)\partial^\rho c\circ A_\rho=0\quad
\Rightarrow\quad a_4=2\xi a_3.
\label{a4val}
\end{equation}
At this point our dictionary reads 
\begin{align}
h_{\mu\nu}&=2A_\mu\circ A_\nu+a_1\eta_{\mu\nu}
\left(A^\rho\circ A_\rho+2\xi c\circ \bar{c}\right)\notag\\
\varphi&=a_3\left(A^\rho\circ A_\rho+2\xi c\circ\bar{c}
\right),
\label{dict2}
\end{align}
and we may fix the remaining coefficients by considering the known
BRST transformations for the level one (anti)-ghosts:
\begin{equation}
Q c_\mu=0,\quad Q\bar{c}_\mu=\frac{1}{\xi}G_\mu[h_{\mu\nu},\varphi], 
\label{antighostBRST}
\end{equation}
where $G_\mu[h_{\mu\nu},\varphi]$ is the gauge-fixing condition in the
gravity theory~\footnote{In principle, we could have introduced a
  second arbitrary multiplier $\xi$ in eq.~(\ref{antighostBRST}), but have
  instead chosen the same $\xi$ as eq.~(\ref{YMBRST}), which does not
  affect our conclusions.}, which may involve both the graviton and
dilaton fields in general. However, we will take it to correspond to
the de Donder gauge:
\begin{equation}
G_\mu=\partial^\nu \bar{h}_{\mu\nu},\quad 
\bar{h}_{\mu\nu}=h_{\mu\nu}-\frac{h}{2}\eta_{\mu\nu},\quad h\equiv 
h^\alpha_\alpha.
\label{deDonder}
\end{equation}
From eq.~(\ref{cresult}) one finds
\begin{align}
Q\bar{c}_\mu&=\frac{2}{\xi}\left[\partial^\rho A_\rho\circ A_\mu
+2\xi\partial_\mu (c\circ \bar{c})\right]\notag\\
&=\frac{1}{\xi}\left[\partial^\rho \bar{h}_{\rho\mu}
+\frac{(1+a_1)}{a_3}\partial_\mu\varphi\right],
\label{Qcbarmu}
\end{align}
where we have used eq.~(\ref{dict2}). Comparing
eqs.~(\ref{antighostBRST}, \ref{deDonder}, \ref{Qcbarmu}), we see that
the only way for the gauge-fixing condition in gravity to correspond
to the de Donder gauge, for arbitrary dilaton fields, is to set
$a_1=-1$.  Furthermore, the only effect of $a_3$ is to rescale the
dilaton field, so we may set $a_3=1$ without loss of generality. Thus,
we have finally obtained the dictionary
\begin{equation}
h_{\mu\nu}=2A_\mu\circ A_\nu-\eta_{\mu\nu}\left(A^\rho\circ A_\rho+2\xi c\circ
\bar{c}\right),\quad 
\varphi=\left(A^\rho\circ A_\rho+2\xi c\circ \bar{c}\right).
\label{dictlocal}
\end{equation}
Let us now apply this to the point charge in pure Yang-Mills theory,
where the Lorenz gauge field is given by~\footnote{The form of the
  Lorenz gauge field is not unique, owing to the possibility of making
  residual gauge transformations that preserve the Lorenz gauge
  condition.}
\begin{equation}
A^a_\mu=\frac{g \alpha^a}{4\pi r}u_\mu,\quad u_\mu=(1,0,0,0),
\label{Amuform}
\end{equation}
where $\alpha^a$ is a colour vector, whose spacetime dependence may be
neglected in the linearised theory (see
e.g. ref.~\cite{Goldberger:2016iau}). This satisfies the equation of
motion
\begin{equation}
\Box A_\mu^a=j^a_\mu,\quad j^a_\mu=g\alpha^a\delta^{(3)}(\vec{x})u_\mu,
\label{AmuEOM}
\end{equation}
where the current indeed represents a point-like colour charge located
at the origin. One must also give the form of the (anti-)ghost fields
$(c,\bar{c})$, which obey analogous equations of motion to
eq.~(\ref{AmuEOM}):
\begin{equation}
\Box c=j,\quad \Box 
\bar{c}=\bar{j},\quad j=gD\delta^{(3)}(\vec{x}),
\quad \bar{j}=g\bar{D}\delta^{(3)}(\vec{x}),
\label{ghostEOM}
\end{equation}
where $(D,\bar{D})$ are constant Grassmann numbers at linear order. It
may at first seem strange that the ghost fields are being classically
sourced, given that in scattering amplitudes they usually only occur
in loops, where their job is to subtract unphysical degrees of
freedom. However, we are here calculating the expectation value of a
classical field, which is an off-shell quantity. It is thus
gauge-variant, and the precise form it takes will depend on the choice
of gauge-fixing, and hence the ghost contributions. Furthermore, we
will see later on that the ghosts have a pivotal role to play when we
take the double copy. Returning to the point charge, the solution of
eq.~(\ref{ghostEOM}) is straightforward:
\begin{equation}
c=\frac{gD}{4\pi r},\quad \bar{c}=\frac{g\bar{D}}{4\pi r},
\label{ghostsol}
\end{equation}
and we can now apply the dictionary of eq.~(\ref{dictlocal}) to double
copy the above results to gravity. We first need the form of the
spectator field that appears in the circle product of
eq.~(\ref{circprod}). This should correspond to the inverse of the
scalar propagator~\cite{Anastasiou:2014qba}, which in the present
(static) case reads~\footnote{Note that we have here introduced a
  factor of the gauge theory coupling $g$ in the spectator field, for
  book-keeping purposes: the double copy dictionary then consists of
  obtaining gravity fields linear in $\kappa$ from gauge fields linear
  in $g$. This has an analogue in the Kerr-Schild double copy of
  ref.~\cite{Monteiro:2014cda}, in which the scalar field used to
  formulate the copy procedure contains a factor of the gauge theory
  coupling constant.}
\begin{equation}
\Phi^{aa'}=\frac{g\delta^{aa'}}{4\pi r}.
\label{Phiform}
\end{equation}
To use eq.~(\ref{dictlocal}), we
may take the Fourier transform ${\cal F}$ to momentum space, such that
the convolutions are replaced by products. Using the simple result
\begin{equation}
{\cal F}\left[\frac{1}{r}\right]=\frac{\delta(k^0)}{\vec{k}^2},
\label{Fourier}
\end{equation}
where $k$ is the 4-momentum variable conjugate to $x$, one finds
\begin{equation}
{\cal F}\left[A_\mu\circ A_\nu\right]=
\frac{g(\alpha\cdot \alpha)\delta(k^0)}{4\pi\vec{k}^2}u_\mu u_\nu,
\quad {\cal F}\left[c\circ\bar{c}\right]=\frac{g (D\cdot \bar{D})
\delta(k^0)}{4\pi\vec{k}^2},
\label{Fourierprods}
\end{equation}
and thus
\begin{equation}
\bar{h}_{\mu\nu}=\frac{g}{4\pi r}\left[2(\alpha\cdot \alpha)u_\mu
  u_\nu +2\xi (D\cdot\bar{D})\eta_{\mu\nu}\right],\quad 
\varphi=\frac{g}{4\pi r}\left[\alpha\cdot\alpha
+2\xi D\cdot \bar{D}\right].
\label{hphiresults}
\end{equation}
Some comments are in order. Firstly, we see that the choice of ghost
sources $(D,\bar{D})$ completely specifies how the trace degree of
freedom of the graviton, and the dilaton field, can be unambiguously
disentangled from each other, at least in principle. This ambiguity
has been noted in other approaches to the classical double
copy~\cite{Luna:2016hge}. Secondly, we may compare
eq.~(\ref{hphiresults}) with the known form of the linearised JNW
solution in the de Donder gauge (see e.g. ref.~\cite{Luna:2016hge}):
\begin{equation}
\bar{h}_{\mu\nu}=\frac{\kappa}{2}\frac{M}{4\pi r}u_\mu u_\nu,\quad
\varphi=-\frac{\kappa}{2}\frac{Y}{4\pi r},
\label{JNWdeDonder}
\end{equation}
where $M$ is the Schwarzschild mass term, and $Y$ an additional
independent parameter. We then immediately see that the only way to
recover the form of eq.~(\ref{JNWdeDonder}) from
eq.~(\ref{hphiresults}) is to choose the ghost sources to vanish
i.e. $D=\bar{D}=0$. In that case, we do not recover the full
two-parameter JNW solution, but only a special case. That is, upon
replacing the coupling constants appropriately, and also the arbitrary
colour charge by the Schwarzschild mass:
\begin{equation}
g\rightarrow\frac{\kappa}{2},\quad 2(\alpha\cdot \alpha)\rightarrow M,
\label{replacements}
\end{equation}
we find that $Y=-M/2$ is fixed, and thus not independent from
$M$. That this is the correct outcome follows directly from the
dictionary of eq.~(\ref{dictlocal}), which implies
\begin{equation}
\bar{h}_{\mu\nu}=2\left[A_\mu\circ A_\nu+\xi\eta_{\mu\nu}c\circ\bar{c}\right].
\label{hbardict}
\end{equation}
We then see directly that one may only achieve the vanishing of the de
Donder gauge constraint of eq.~(\ref{deDonder}) upon imposing the Lorenz constraint on the gauge field  by setting the
(anti-)ghost fields to zero.\\

Some care is needed to interpret what has gone on here. It would not
be correct, for example, to conclude that the double copy of the point
charge is only a special case of JNW: this would directly contradict
the results of refs.~\cite{Monteiro:2014cda,Kim:2019jwm}. Instead the
problem can be traced back to the fact that we have demanded that the
double copy dictionary of eqs.~(\ref{dict1a}, \ref{dict1b}) did not
contain non-local derivative operators in position space. This choice
has proven to be overly restrictive, and indeed insufficient to
explore the full double copy of the point charge. As we will see in
the following section, a very different situation occurs if the
locality assumption is relaxed. However, it should also be stressed
that the results of this section are also specific to having chosen
the Lorenz and de Donder gauges in the gauge and gravity theory
respectively - it may well be that a local dictionary can lead to more
general JNW solutions if a different gauge is chosen.

\section{The JNW solution from a general dictionary}
\label{sec:nonlocal}

The results of the previous section suggest that a more general
dictionary is needed in order to relate gauge and graviton fields in
the Lorenz and de Donder gauges. To this end, we may write the most
general possible covariant ansatz for the graviton and dilaton containing gauge,
(anti-)ghost fields and (inverse) derivative operators, as follows:
\begin{align}
h_{\mu\nu}&=2A_\mu\circ A_\nu+\frac{d_1}{\Box}\partial_{(\mu}A_{\nu)}
\circ \partial A+\frac{d_2}{\Box^2}\partial_\mu\partial_\nu \partial A
\circ \partial A+d_3\frac{\partial_\mu\partial_\nu}{\Box}
A^\rho\circ A_\rho + d_4\frac{\partial_\mu\partial_\nu}{\Box} c\circ\bar{c}
\notag\\
&\quad+\eta_{\mu\nu}\left[a_1 A^\rho\circ A_\rho +\frac{a_2}{\Box}\partial A
\circ \partial A+a_3 c\circ\bar{c}\right];\notag\\
\varphi&=b_1 A^\rho\circ A_\rho+\frac{b_2}{\Box}\partial A\circ \partial A
+a_3 c\circ\bar{c},
\label{generaldict}
\end{align}
where we use the slight shorthand notation $\partial A\equiv
\partial\cdot A$. As in the previous section, imposing the BRST
transformations of eq.~(\ref{gravBRST}) leads to constraints from the
coefficient of $\eta_{\mu\nu}$ in the graviton, and also the dilaton,
which in this case turn out to be
\begin{equation}
a_3=2\xi(a_1+a_2),\quad b_3=2\xi (b_1+b_2).
\label{a3b3}
\end{equation}
One also finds expressions for the first-level gravitational
ghost:
\begin{align}
c_\mu&=\left(2+\frac{d_1}{2}\right)A_\mu\circ c+\left(\frac{d_1}{2}
+d_2+d_3-\frac{d_4}{2\xi}\right)\frac{\partial_\mu}{\Box}(c\circ \partial A),
\label{ghostsgeneral}
\end{align}
where $\bar{c}_\mu$ is obtained by conjugation by replacing
$c\rightarrow\bar{c}$. We can then impose the anti-ghost
transformation of eq.~(\ref{antighostBRST}) on the latter, which
yields
\begin{align}
\frac{1}{\xi}G_\mu[h]&=\left(2+\frac{d_1}{2}\right)\frac{1}{\xi}A_\mu\circ \partial A
+\left(\frac{d_1}{2}+d_2+d_3-\frac{d_4}{2\xi}\right)\frac{\partial_\mu}{\Box}
\partial A\circ \partial A\notag\\
&\quad+\left(2+d_1+d_2+d_3-\frac{d_4}{2\xi}\right)
\partial_\mu(c\circ\bar{c}).
\label{Qbarc}
\end{align}
However, we want the gauge-fixing condition in gravity to vanish when
the Lorenz gauge condition $\partial A=0$ is satisfied in gauge
theory~\footnote{In previous works on the convolutional double copy,
  the BRST gauge mapping was enforced without requiring the
  gauge-fixing condition in gravity to vanish if its gauge theory
  counterpart does. Here we need to implement such conditions, given
  that we are considering explicit solutions.}, immediately implying
\begin{equation}
d_4=2\xi(2+d_1+d_2+d_3),
\label{d4}
\end{equation}
which we may use to tidy up eq.~(\ref{Qbarc}):
\begin{equation}
G_\mu[h]=\left(2+\frac{d_1}{2}\right)\left[A_\mu\circ\partial A
-\frac{\partial_\mu}{\Box}\partial A\circ \partial A\right].
\label{Gmures}
\end{equation}
Furthermore, we may substitute the constraints of eqs.~(\ref{a3b3},
\ref{d4}) into eq.~(\ref{generaldict}) and derive, after some work,
\begin{align}
\partial^\nu\bar{h}_{\mu\nu}&=\left(2+\frac{d_1}{2}\right)\partial A\circ A_\mu
-\left(a_2-\frac{d_2}{2}\right)\frac{\partial_\mu}{\Box}\partial A\circ 
\partial A-\left(1+a_1-\frac{d_3}{2}\right)\partial_\mu A^\rho\circ A_\rho
\notag\\
&\quad
-2\xi\left(-1+a_1+a_2-\frac{d_1+d_2+d_3}{2}\right)\partial_\mu c\circ\bar{c}.
\label{dhmunu}
\end{align}
Equating this with eq.~(\ref{Gmures}) yields the additional
constraints
\begin{equation}
a_2=2+\frac{d_1+d_2}{2},\quad a_1=-1+\frac{d_3}{2},
\label{a2a1}
\end{equation}
such that substituting these into eq.~(\ref{generaldict}) yields the
most general dictionary relating the Lorenz and de Donder fields:
\begin{align}
h_{\mu\nu}&=2A_\mu\circ A_\nu+\frac{d_1}{\Box}\partial_{(\mu}A_{\nu)}\circ
\partial A+\frac{d_2}{\Box^2}\partial_\mu\partial_\nu\partial A\circ 
\partial A+d_3\frac{\partial_\mu\partial_\nu}{\Box}A^\rho\circ A_\rho
\notag\\
&+2\xi(2+d_1+d_2+d_3)\frac{\partial_\mu\partial_\nu}{\Box}c\circ\bar{c}
+\eta_{\mu\nu}\left[\left(-1+\frac{d_3}{2}\right)A^\rho\circ A_\rho
+\left(2+\frac{d_1+d_2}{2}\right)\frac{1}{\Box}\partial A\circ\partial A
\right.\notag\\
&\left.\quad+2\xi\left(1+\frac{d_1+d_2+d_3}{2}\right)
c\circ\bar{c}\right];\notag\\
\varphi&=b_1 A^\rho\circ A_\rho+\frac{b_2}{\Box}\partial A\circ \partial A
+2\xi(b_1+b_2)c\circ\bar{c}.
\label{generaldict2}
\end{align}
We thus see that the requirement that our two gauge choices match is
insufficient to completely fix the dictionary, leaving a number of
remaining arbitrary parameters. These represent the ability to make
residual gauge transformations and field redefinitions that preserve
the de Donder gauge condition. To examine the double copy of the point
charge, it is sufficient to choose a particular case of the
dictionary, and to this end we will choose
\begin{equation}
d_1=-2,\quad d_2=d_3=0,\quad b_1=2,\quad b_2=0,
\label{parameterchoices}
\end{equation}
leading to the simple dictionary
\begin{equation}
\bar{h}_{\mu\nu}=2A_\mu\circ A_\nu-\frac{2}{\Box}\partial_{(\mu}A_{\nu)}
\circ \partial A,\quad \varphi=2A^\rho\circ A_\rho +4\xi c\circ\bar{c}.
\label{simpledict}
\end{equation}
Fourier transforming to momentum space and using the results of
eqs.~(\ref{Amuform}, \ref{ghostsol}, \ref{Phiform}), one now finds
that the graviton and dilaton fields are given by
\begin{equation}
\bar{h}_{\mu\nu}=\frac{g(\alpha\cdot\alpha)}{4\pi r}u_\mu u_\nu,\quad
\phi=\frac{g}{4\pi r}\left[(\alpha\cdot\alpha)+4\xi D\cdot \bar{D}\right].
\label{hphigeneral}
\end{equation}
We can recognise this as the full two-parameter JNW solution, provided
that we identify
\begin{equation}
g\rightarrow \frac{\kappa}{2},\quad (\alpha\cdot\alpha)\rightarrow M,\quad 
[(\alpha\cdot\alpha)+4g\xi D\cdot\bar{D}]\rightarrow -Y.
\label{JNWreplace}
\end{equation}
The first and second of these replacements correspond to the usual
replacement of couplings in the double copy, and colour charge by
Schwarzschild mass~\cite{Monteiro:2014cda}. The third replacement
generates the coefficient of the dilaton in the JNW solution, which is
then determined by the ghost sources. Thus, we see that the full
two-parameter JNW solution is obtained, commensurate with
refs.~\cite{Luna:2016hge,Kim:2019jwm}. Furthermore, the convolutional
double copy provides a key insight into where these two parameters
come from, namely the ability to source the ghosts and gauge fields
(whilst keeping the Lorenz and de Donder gauges)
independently. Indeed, the gravitational equations of
motion~\footnote{We have here chosen $\xi=-1$ for simplicity, which
  does not affect our argument.}
\begin{equation}
\frac12 \Box \bar{h}_{\mu\nu}=j_{\mu\nu},\quad \Box \varphi=j_\varphi,
\label{EOMgravity}
\end{equation}
together with eqs.~(\ref{AmuEOM}, \ref{ghostEOM}) (with sources left
general), and the dictionary of eq.~(\ref{simpledict}), imply the
following relations between the sources in the gauge and gravity
theory:
\begin{align}
j_{\mu\nu}&=\frac{1}{\Box}j_\mu\circ
j_\nu-\frac{1}{\Box^2}\partial_{(\mu} j_{\nu)}\circ \partial j,\notag\\
j_\varphi&=\frac{2}{\Box}j^\rho\circ j_\rho-\frac{4}{\Box}j\circ\bar{j}.
\label{sourcedict}
\end{align}
These relations directly encode the fact that independence of the
graviton and dilaton sources are inherited from the separate gauge and
(anti-)ghost sources in the single copy theory.\\

It is instructive to compare our results with those of
ref.~\cite{Luna:2016hge}, which also considered how to obtain the JNW
solution from the double copy. By applying a generalisation of the BCJ
double copy to perturbative solutions of the equation of motion, the
authors defined a {\it fat graviton field} $H_{\mu\nu}$, satisfying
the linearised equation of motion
\begin{equation}
\Box H_{\mu\nu}=0,
\label{HEOM}
\end{equation}
and which combines the physical degrees of freedom of the graviton,
dilaton and axion. One must then provide a prescription for obtaining
the individual fields (in a particular generalised gauge) from the fat
graviton. To this end, ref.~\cite{Luna:2016hge} used the ansatz (at
linearised level)
\begin{equation}
H_{\mu\nu}=\bar{h}_{\mu\nu}+B_{\mu\nu}+P_{\mu\nu}^q(\varphi-\bar{h}),
\label{Hdecomp}
\end{equation}
where $\bar{h}=\bar{h}^\mu_\mu$, and we have defined the projection
operator
\begin{equation}
P_{\mu\nu}^q=\frac{1}{d-2}\left(\eta_{\mu\nu}+\frac{q_\mu\partial_\nu
+q_\nu\partial_\mu}{q\cdot \partial}\right),
\label{Pdef}
\end{equation}
where $q$ is a constant null 4-vector such that $q\cdot k\neq 0$, if
$k$ is the momentum of the fat graviton. The individual fields
(referred to as {\it skinny fields} in ref.~\cite{Luna:2016hge}) are
given by
\begin{align}
\bar{h}_{\mu\nu}&=\frac12\left(H_{\mu\nu}+H_{\nu\mu}\right)-P_{\mu\nu}^qH,
\notag\\
B_{\mu\nu}&=\frac12\left(H_{\mu\nu}-H_{\nu\mu}\right),\notag\\
\varphi&=H,
\label{skinnyfields}
\end{align}
where one may show that the graviton thus obtained is in de Donder
gauge. The role of the projection operator of eq.~(\ref{Pdef}) is to
resolve the ambiguity regarding how to divide the trace of the fat
graviton ($H$) between the graviton trace ($\bar{h}$) and dilaton
($\varphi$) degrees of freedom. In eq.~(\ref{skinnyfields}), all of
$H$ is taken to correspond to the dilaton, although it is possible to
mix up $\varphi$ and $\bar{h}$ by performing residual gauge
transformations that preserve the de Donder gauge condition. Using
this approach, ref.~\cite{Luna:2016hge} showed that the most general
fat graviton for the JNW solution can be written as
\begin{equation}
H_{\mu\nu}=\frac{\kappa}{2}\frac{1}{4\pi r}
\left(M u_\mu u_\nu+(M-Y)\frac12\left(\eta_{\mu\nu}
-q_\mu l_\nu-q_\nu l_\mu\right)\right),
\label{HJNW}
\end{equation}
where $u_\mu$ has been defined above, amd $l_\mu$ is a 4-vector such
that $q\cdot l=1$. This reproduces the de Donder gauge graviton and
dilaton fields of eq.~(\ref{JNWdeDonder}), and is furthermore such
that the single copy is always a gauge theory point
charge. Furthermore, ref.~\cite{Luna:2016hge} showed how
eq.~(\ref{HJNW}) could be used as a building block in understanding
the classical double copy at higher perturbative orders.\\

It is interesting to note a number of similarities between the
convolutional double copy and fat graviton approaches. Firstly, they
both produce the full two-parameter JNW solution in the gravity
theory. Secondly, there is an ambiguity in how one obtains the
latter. In the convolutional approach, this manifests itself in the
free parameters entering the dictionary for the graviton and dilaton
fields, once the relevant BRST transformations for the Lorenz and de
Donder gauge conditions have already been imposed. In
eq.~(\ref{HJNW}), the ambiguity appears as the arbitrary vector $q$.
Different choices of this vector - together with residual gauge
transformations preserving the de Donder gauge condition - can be used
to map out the full parameter space $(M,Y)$ of the JNW solution, by
deciding whether or not one sources the dilaton. In the convolutional
approach, this is the job played by the BRST ghosts, and both
approaches have something in common with the double copy for
scattering amplitudes, in which the external degrees of freedom
(graviton, dilaton or axion) appearing in a gravity amplitude depend
on the physical polarisation states one chooses in the two gauge
theory amplitudes entering the double copy. \\

A third similarity between the convolutional and fat graviton
approaches concerns the use of non-local operators in position
space. In eq.~(\ref{skinnyfields}), the projector $P_{\mu\nu}^q$ is
non-local, as can be seen from its definition in
eq.~(\ref{Pdef}). This mirrors the use of non-local terms in the
convolutional dictionary of eq.~(\ref{generaldict}), in order to be
able to obtain the full JNW solution rather than a special case. There
may, of course, be choices of gauge in both approaches such that
non-local terms are not needed, but it is nevertheless intriguing that
non-locality arises whatever method one chooses for performing the
double copy to a de Donder gauge solution.

\section{Conclusion}
\label{sec:conclude}

In this paper, we have examined a particular approach for obtaining
linearised solutions of gravity theories as a double copy of gauge
theory, namely the convolutional approach of
refs.~\cite{Anastasiou:2014qba,Anastasiou:2016csv,Cardoso:2016ngt,Cardoso:2016amd,Anastasiou:2017nsz,Anastasiou:2017taf,LopesCardoso:2018xes,Anastasiou:2018rdx,Borsten:2019prq,Borsten:2020bgv}. Our
aim was twofold. Firstly, it is instructive to clarify how to use this
approach by looking at particular solutions. Secondly, by choosing the
point charge in pure Yang-Mills theory, the convolutional approach
offers a useful complementary point of view to related approaches,
that have debated whether the full JNW solution in ${\cal N}=0$
supergravity - or a special case of this - is the appropriate gravity
solution~\cite{Monteiro:2014cda,Goldberger:2017frp,Luna:2016hge,Kim:2019jwm}.\\

To use the convolutional double copy, one must posit a suitable
dictionary between the gauge and gravity fields, containing free
parameters. The latter can then be at least partially fixed by
imposing the correct BRST transformations in both theories, which
explicitly brings in the gauge-fixing conditions. Unfixed parameters
correspond to residual generalised gauge transformations that leave
the gauge-fixing condition unchanged. In examining the point charge in
the Lorenz gauge, we found that a completely local dictionary was
insufficient to obtain the full two-parameter JNW solution in the
gravity theory in de Donder gauge, as seen in the previous results of
refs.~\cite{Luna:2016hge,Kim:2019jwm}. The solution to this problem
was to include non-local operators in position space, which indeed
allowed us to obtain the full JNW solution. An arbitrary combination
of $M$ and $Y$ in this solution can be made by choosing sources for
the BRST ghost fields appropriately, mirroring the role of the
arbitrary gauge-vectors in ref.~\cite{Luna:2016hge}. Furthermore, many
free parameters in the general dictionary remained unconstrained,
indicating that the double copy to de Donder gauge is robust under a
non-trivial group of generalised gauge transformations.\\

There are many possible avenues for further work. It would be
interesting, for example, to relate the convolutional approach to the
Kerr-Schild approach of ref.~\cite{Monteiro:2014cda}. Although this
seems a natural thing to do - given that the Kerr-Schild procedure
linearises the field equations on both sides of the double copy - it
is not immediately clear how to proceed. Not all metrics, for example,
can be expressed in Kerr-Schild coordinates, so that there is no such
thing as a general ``Kerr-Schild gauge''. Without an explicit
gauge-fixing term, it is difficult to unambiguously apply the BRST
procedure to relate the gauge and gravity theories. Another important
development would be to extend the convolutional double copy in the
BRST context to non-linear orders in perturbation theory. This could
provide key insights into obtaining pure gravity perturbation theory
from the double copy, given that the convolution approach makes clear
how the dilaton is generated (or not) in the gravity theory, and can
also be used in arbitrary gauges. Work in these and other directions
is in progress~\cite{Borsten_Nagy}.


\section*{Acknowledgments}

We thank Leron Borsten and Donal O'Connell for useful conversations. This work has been
supported by the UK Science and Technology Facilities Council (STFC)
Consolidated Grant ST/P000754/1 ``String theory, gauge theory and
duality'', and by the European Union Horizon 2020 research and
innovation programme under the Marie Sk\l{}odowska-Curie grant
agreement No. 764850 ``SAGEX''. AL is supported in part by the Department of Energy under Award Number DESC000993. SN is supported by STFC grant ST/P000703/1 and a Leverhulme Research Project Grant. This research was supported by the
Munich Institute for Astro- and Particle Physics (MIAPP) which is
funded by the Deutsche Forschungsgemeinschaft (DFG, German Research
Foundation) under Germany´s Excellence Strategy – EXC-2094 – 390783311.

\bibliography{refs}

\providecommand{\href}[2]{#2}\begingroup\raggedright\begin{thebibliography}{10}

\bibitem{Bern:2008qj}
Z.~Bern, J.~Carrasco, and H.~Johansson, ``{New Relations for Gauge-Theory
  Amplitudes},'' {\em Phys.Rev.} {\bf D78} (2008) 085011,
\href{http://www.arXiv.org/abs/0805.3993}{{\tt 0805.3993}}.

\bibitem{Bern:2010ue}
Z.~Bern, J.~J.~M. Carrasco, and H.~Johansson, ``{Perturbative Quantum Gravity
  as a Double Copy of Gauge Theory},'' {\em Phys.Rev.Lett.} {\bf 105} (2010)
  061602, \href{http://www.arXiv.org/abs/1004.0476}{{\tt 1004.0476}}.

\bibitem{Bern:2010yg}
Z.~Bern, T.~Dennen, Y.-t. Huang, and M.~Kiermaier, ``{Gravity as the Square of
  Gauge Theory},'' {\em Phys.Rev.} {\bf D82} (2010) 065003,
  \href{http://www.arXiv.org/abs/1004.0693}{{\tt 1004.0693}}.

\bibitem{Monteiro:2014cda}
R.~Monteiro, D.~O'Connell, and C.~D. White, ``{Black holes and the double
  copy},'' {\em JHEP} {\bf 1412} (2014) 056,
\href{http://www.arXiv.org/abs/1410.0239}{{\tt 1410.0239}}.

\bibitem{Luna:2015paa}
A.~Luna, R.~Monteiro, D.~O'Connell, and C.~D. White, ``{The classical double
  copy for Taub-NUT spacetime},'' {\em Phys. Lett.} {\bf B750} (2015) 272--277,
\href{http://www.arXiv.org/abs/1507.01869}{{\tt 1507.01869}}.

\bibitem{Luna:2016due}
A.~Luna, R.~Monteiro, I.~Nicholson, D.~O'Connell, and C.~D. White, ``{The
  double copy: Bremsstrahlung and accelerating black holes},''
\href{http://www.arXiv.org/abs/1603.05737}{{\tt 1603.05737}}.

\bibitem{Bahjat-Abbas:2017htu}
N.~Bahjat-Abbas, A.~Luna, and C.~D. White, ``{The Kerr-Schild double copy in
  curved spacetime},'' {\em JHEP} {\bf 12} (2017) 004,
\href{http://www.arXiv.org/abs/1710.01953}{{\tt 1710.01953}}.

\bibitem{Berman:2018hwd}
D.~S. Berman, E.~Chacón, A.~Luna, and C.~D. White, ``{The self-dual classical
  double copy, and the Eguchi-Hanson instanton},''
\href{http://www.arXiv.org/abs/1809.04063}{{\tt 1809.04063}}.

\bibitem{Ridgway:2015fdl}
A.~K. Ridgway and M.~B. Wise, ``{Static Spherically Symmetric Kerr-Schild
  Metrics and Implications for the Classical Double Copy},'' {\em Phys. Rev.}
  {\bf D94} (2016), no.~4, 044023,
\href{http://www.arXiv.org/abs/1512.02243}{{\tt 1512.02243}}.

\bibitem{Carrillo-Gonzalez:2017iyj}
M.~Carrillo-González, R.~Penco, and M.~Trodden, ``{The classical double copy
  in maximally symmetric spacetimes},'' {\em JHEP} {\bf 04} (2018) 028,
\href{http://www.arXiv.org/abs/1711.01296}{{\tt 1711.01296}}.

\bibitem{CarrilloGonzalez:2019gof}
M.~Carrillo~González, B.~Melcher, K.~Ratliff, S.~Watson, and C.~D. White,
  ``{The classical double copy in three spacetime dimensions},'' {\em JHEP}
  {\bf 07} (2019) 167,
\href{http://www.arXiv.org/abs/1904.11001}{{\tt 1904.11001}}.

\bibitem{Bahjat-Abbas:2020cyb}
N.~Bahjat-Abbas, R.~Stark-Muchão, and C.~D. White, ``{Monopoles, shockwaves
  and the classical double copy},''
\href{http://www.arXiv.org/abs/2001.09918}{{\tt 2001.09918}}.

\bibitem{Alfonsi:2020lub}
L.~Alfonsi, C.~D. White, and S.~Wikeley, ``{Topology and Wilson lines: global
  aspects of the double copy},''
  \href{http://www.arXiv.org/abs/2004.07181}{{\tt 2004.07181}}.

\bibitem{Alawadhi:2019urr}
R.~Alawadhi, D.~S. Berman, B.~Spence, and D.~Peinador~Veiga, ``{S-duality and
  the double copy},'' {\em JHEP} {\bf 03} (2020) 059,
  \href{http://www.arXiv.org/abs/1911.06797}{{\tt 1911.06797}}.

\bibitem{Monteiro:2018xev}
R.~Monteiro, I.~Nicholson, and D.~O'Connell, ``{Spinor-helicity and the
  algebraic classification of higher-dimensional spacetimes},''
\href{http://www.arXiv.org/abs/1809.03906}{{\tt 1809.03906}}.

\bibitem{Luna:2018dpt}
A.~Luna, R.~Monteiro, I.~Nicholson, and D.~O'Connell, ``{Type D Spacetimes and
  the Weyl Double Copy},'' {\em Class. Quant. Grav.} {\bf 36} (2019) 065003,
\href{http://www.arXiv.org/abs/1810.08183}{{\tt 1810.08183}}.

\bibitem{Goldberger:2016iau}
W.~D. Goldberger and A.~K. Ridgway, ``{Radiation and the classical double copy
  for color charges},'' {\em Phys. Rev.} {\bf D95} (2017), no.~12, 125010,
\href{http://www.arXiv.org/abs/1611.03493}{{\tt 1611.03493}}.

\bibitem{Goldberger:2017frp}
W.~D. Goldberger, S.~G. Prabhu, and J.~O. Thompson, ``{Classical gluon and
  graviton radiation from the bi-adjoint scalar double copy},'' {\em Phys.
  Rev.} {\bf D96} (2017), no.~6, 065009,
\href{http://www.arXiv.org/abs/1705.09263}{{\tt 1705.09263}}.

\bibitem{Goldberger:2017vcg}
W.~D. Goldberger and A.~K. Ridgway, ``{Bound states and the classical double
  copy},'' {\em Phys. Rev.} {\bf D97} (2018), no.~8, 085019,
\href{http://www.arXiv.org/abs/1711.09493}{{\tt 1711.09493}}.

\bibitem{Goldberger:2017ogt}
W.~D. Goldberger, J.~Li, and S.~G. Prabhu, ``{Spinning particles, axion
  radiation, and the classical double copy},'' {\em Phys. Rev.} {\bf D97}
  (2018), no.~10, 105018,
\href{http://www.arXiv.org/abs/1712.09250}{{\tt 1712.09250}}.

\bibitem{Goldberger:2019xef}
W.~D. Goldberger and J.~Li, ``{Strings, extended objects, and the classical
  double copy},''
\href{http://www.arXiv.org/abs/1912.01650}{{\tt 1912.01650}}.

\bibitem{Luna:2016hge}
A.~Luna, R.~Monteiro, I.~Nicholson, A.~Ochirov, D.~O'Connell, N.~Westerberg,
  and C.~D. White, ``{Perturbative spacetimes from Yang-Mills theory},'' {\em
  JHEP} {\bf 04} (2017) 069,
\href{http://www.arXiv.org/abs/1611.07508}{{\tt 1611.07508}}.

\bibitem{Luna:2017dtq}
A.~Luna, I.~Nicholson, D.~O'Connell, and C.~D. White, ``{Inelastic Black Hole
  Scattering from Charged Scalar Amplitudes},'' {\em JHEP} {\bf 03} (2018) 044,
\href{http://www.arXiv.org/abs/1711.03901}{{\tt 1711.03901}}.

\bibitem{Maybee:2019jus}
B.~Maybee, D.~O'Connell, and J.~Vines, ``{Observables and amplitudes for
  spinning particles and black holes},''
\href{http://www.arXiv.org/abs/1906.09260}{{\tt 1906.09260}}.

\bibitem{Borsten_Nagy}
L.~Borsten and S.~Nagy, ``{The double-copy to cubic order (with ghosts)},''
  {\em to appear}.

\bibitem{Lee:2018gxc}
K.~Lee, ``{Kerr-Schild Double Field Theory and Classical Double Copy},''
\href{http://www.arXiv.org/abs/1807.08443}{{\tt 1807.08443}}.

\bibitem{Kim:2019jwm}
K.~Kim, K.~Lee, R.~Monteiro, I.~Nicholson, and D.~Peinador~Veiga, ``{The
  Classical Double Copy of a Point Charge},''
\href{http://www.arXiv.org/abs/1912.02177}{{\tt 1912.02177}}.

\bibitem{Anastasiou:2014qba}
A.~Anastasiou, L.~Borsten, M.~J. Duff, L.~J. Hughes, and S.~Nagy, ``{Yang-Mills
  origin of gravitational symmetries},'' {\em Phys. Rev. Lett.} {\bf 113}
  (2014), no.~23, 231606,
\href{http://www.arXiv.org/abs/1408.4434}{{\tt 1408.4434}}.

\bibitem{Anastasiou:2016csv}
A.~Anastasiou, L.~Borsten, M.~J. Duff, M.~J. Hughes, A.~Marrani, S.~Nagy, and
  M.~Zoccali, ``{Twin supergravities from Yang-Mills theory squared},'' {\em
  Phys. Rev.} {\bf D96} (2017), no.~2, 026013,
\href{http://www.arXiv.org/abs/1610.07192}{{\tt 1610.07192}}.

\bibitem{Cardoso:2016ngt}
G.~L. Cardoso, S.~Nagy, and S.~Nampuri, ``{A double copy for $ \mathcal{N}=2 $
  supergravity: a linearised tale told on-shell},'' {\em JHEP} {\bf 10} (2016)
  127,
\href{http://www.arXiv.org/abs/1609.05022}{{\tt 1609.05022}}.

\bibitem{Cardoso:2016amd}
G.~Cardoso, S.~Nagy, and S.~Nampuri, ``{Multi-centered $ \mathcal{N}=2 $ BPS
  black holes: a double copy description},'' {\em JHEP} {\bf 04} (2017) 037,
\href{http://www.arXiv.org/abs/1611.04409}{{\tt 1611.04409}}.

\bibitem{Anastasiou:2017nsz}
A.~Anastasiou, L.~Borsten, M.~J. Duff, A.~Marrani, S.~Nagy, and M.~Zoccali,
  ``{Are all supergravity theories Yang-Mills squared?},''
\href{http://www.arXiv.org/abs/1707.03234}{{\tt 1707.03234}}.

\bibitem{Anastasiou:2017taf}
A.~Anastasiou, L.~Borsten, M.~J. Duff, A.~Marrani, S.~Nagy, and M.~Zoccali,
  ``{The Mile High Magic Pyramid},''
\newblock 2017.
\newblock
\href{http://www.arXiv.org/abs/1711.08476}{{\tt 1711.08476}}.
\newblock

\bibitem{Borsten:2013bp}
L.~Borsten, M.~Duff, L.~Hughes, and S.~Nagy, ``{Magic Square from Yang-Mills
  Squared},'' {\em Phys. Rev. Lett.} {\bf 112} (2014), no.~13, 131601,
  \href{http://www.arXiv.org/abs/1301.4176}{{\tt 1301.4176}}.

\bibitem{Anastasiou:2013hba}
A.~Anastasiou, L.~Borsten, M.~Duff, L.~Hughes, and S.~Nagy, ``{A magic pyramid
  of supergravities},'' {\em JHEP} {\bf 04} (2014) 178,
  \href{http://www.arXiv.org/abs/1312.6523}{{\tt 1312.6523}}.

\bibitem{Anastasiou:2015vba}
A.~Anastasiou, L.~Borsten, M.~Hughes, and S.~Nagy, ``{Global symmetries of
  Yang-Mills squared in various dimensions},''
\href{http://www.arXiv.org/abs/1502.05359}{{\tt 1502.05359}}.

\bibitem{Bern:2017yxu}
Z.~Bern, J.~J. Carrasco, W.-M. Chen, H.~Johansson, and R.~Roiban, ``{Gravity
  Amplitudes as Generalized Double Copies of Gauge-Theory Amplitudes},'' {\em
  Phys. Rev. Lett.} {\bf 118} (2017), no.~18, 181602,
\href{http://www.arXiv.org/abs/1701.02519}{{\tt 1701.02519}}.

\bibitem{Bern:2017ucb}
Z.~Bern, J.~J.~M. Carrasco, W.-M. Chen, H.~Johansson, R.~Roiban, and M.~Zeng,
  ``{Five-loop four-point integrand of $N=8$ supergravity as a generalized
  double copy},'' {\em Phys. Rev.} {\bf D96} (2017), no.~12, 126012,
\href{http://www.arXiv.org/abs/1708.06807}{{\tt 1708.06807}}.

\bibitem{Faddeev:1967fc}
L.~D. Faddeev and V.~N. Popov, ``{Feynman Diagrams for the Yang-Mills Field},''
  {\em Phys. Lett.} {\bf 25B} (1967)
29--30.

\bibitem{Becchi:1974xu}
C.~Becchi, A.~Rouet, and R.~Stora, ``{The Abelian Higgs-Kibble Model. Unitarity
  of the S Operator},'' {\em Phys. Lett.} {\bf 52B} (1974)
344--346.

\bibitem{Becchi:1974md}
C.~Becchi, A.~Rouet, and R.~Stora, ``{Renormalization of the Abelian
  Higgs-Kibble Model},'' {\em Commun. Math. Phys.} {\bf 42} (1975)
127--162.

\bibitem{Becchi:1975nq}
C.~Becchi, A.~Rouet, and R.~Stora, ``{Renormalization of Gauge Theories},''
  {\em Annals Phys.} {\bf 98} (1976)
287--321.

\bibitem{Tyutin:1975qk}
I.~V. Tyutin, ``{Gauge Invariance in Field Theory and Statistical Physics in
  Operator Formalism},''
\href{http://www.arXiv.org/abs/0812.0580}{{\tt 0812.0580}}.

\bibitem{Fradkin:1975cq}
E.~S. Fradkin and G.~A. Vilkovisky, ``{QUANTIZATION OF RELATIVISTIC SYSTEMS
  WITH CONSTRAINTS},'' {\em Phys. Lett.} {\bf 55B} (1975)
224--226.

\bibitem{Batalin:1977pb}
I.~A. Batalin and G.~A. Vilkovisky, ``{Relativistic S Matrix of Dynamical
  Systems with Boson and Fermion Constraints},'' {\em Phys. Lett.} {\bf 69B}
  (1977)
309--312.

\bibitem{Anastasiou:2018rdx}
A.~Anastasiou, L.~Borsten, M.~J. Duff, S.~Nagy, and M.~Zoccali, ``{Gravity as
  Gauge Theory Squared: A Ghost Story},'' {\em Phys. Rev. Lett.} {\bf 121}
  (2018), no.~21, 211601,
\href{http://www.arXiv.org/abs/1807.02486}{{\tt 1807.02486}}.

\bibitem{Borsten:2019prq}
L.~Borsten, I.~Jubb, V.~Makwana, and S.~Nagy, ``{Gauge $\times$ Gauge on
  Spheres},''
\href{http://www.arXiv.org/abs/1911.12324}{{\tt 1911.12324}}.

\bibitem{Siegel:1988qu}
W.~Siegel, ``{SUPERSTRINGS GIVE OLD MINIMAL SUPERGRAVITY},'' {\em Phys. Lett.}
  {\bf B211} (1988)
55--58.

\bibitem{Siegel:1995px}
W.~Siegel, ``{Curved extended superspace from Yang-Mills theory a la
  strings},'' {\em Phys. Rev.} {\bf D53} (1996) 3324--3336,
\href{http://www.arXiv.org/abs/hep-th/9510150}{{\tt hep-th/9510150}}.

\bibitem{LopesCardoso:2018xes}
G.~Lopes~Cardoso, G.~Inverso, S.~Nagy, and S.~Nampuri, ``{Comments on the
  double copy construction for gravitational theories},'' in {\em {17th
  Hellenic School and Workshops on Elementary Particle Physics and Gravity
  (CORFU2017) Corfu, Greece, September 2-28, 2017}}.
\newblock 2018.
\newblock
\href{http://www.arXiv.org/abs/1803.07670}{{\tt 1803.07670}}.
\newblock

\bibitem{Borsten:2020bgv}
L.~Borsten,
``{Gravity as the square of gauge theory: a review},''.

\bibitem{Janis:1968zz}
A.~I. Janis, E.~T. Newman, and J.~Winicour, ``{Reality of the Schwarzschild
  Singularity},'' {\em Phys. Rev. Lett.} {\bf 20} (1968)
878--880.

\bibitem{White:2016jzc}
C.~D. White, ``{Exact solutions for the biadjoint scalar field},'' {\em Phys.
  Lett.} {\bf B763} (2016) 365--369,
\href{http://www.arXiv.org/abs/1606.04724}{{\tt 1606.04724}}.

\bibitem{DeSmet:2017rve}
P.-J. De~Smet and C.~D. White, ``{Extended solutions for the biadjoint scalar
  field},'' {\em Phys. Lett.} {\bf B775} (2017) 163--167,
\href{http://www.arXiv.org/abs/1708.01103}{{\tt 1708.01103}}.

\bibitem{Bahjat-Abbas:2018vgo}
N.~Bahjat-Abbas, R.~Stark-Muchão, and C.~D. White, ``{Biadjoint wires},'' {\em
  Phys. Lett.} {\bf B788} (2019) 274--279,
\href{http://www.arXiv.org/abs/1810.08118}{{\tt 1810.08118}}.

\end{thebibliography}\endgroup
\end{document}